\documentclass[prd,preprint,preprintnumbers,nofootinbib,eqsecnum,superscriptaddress]{revtex4}

 \usepackage[dvips,final]{graphicx}
  \usepackage{amssymb}
   \usepackage{amsmath}
    \usepackage{amsfonts}
     \usepackage{epsfig}
      \usepackage{bm}% bold math

\usepackage{mathpazo}
\usepackage{ulem}
\usepackage[section]{placeins}

\usepackage{booktabs}
\usepackage{array}
\usepackage{tabularx}
\usepackage{xcolor}
\usepackage{pstricks}
\definecolor{fred}{rgb}{0.90053, 0.00369, 0.00159}  % ta3skyblue

\newcommand{\bp}{\mbox{$\vec p_T$}}
\newcommand{\bq}{\mbox{$\vec q_T$}}

\newcommand{\bk}{\mbox{$\vec k_T$}}

\newcommand{\ket}[1]{ {#1} \rangle}

\newcommand{\half}{{1\over 2}}

\begin{document}
\title{Structure and production mechanism of the enigmatic $X(3872)$ in high-energy hadronic reactions}

\author{Anna Cisek}
\email{acisek@ur.edu.pl}
\affiliation{College of Natural Sciences, Institute of Physics, University of Rzesz\'ow,
ul. Pigonia 1, PL-35-310 Rzesz\'ow, Poland}

\author{Wolfgang Sch\"afer}
\email{Wolfgang.Schafer@ifj.edu.pl}
\affiliation{Institute of Nuclear Physics, Polish Academy of Sciences,
ul. Radzikowskiego 152, PL-31-342 Krak{\'o}w, Poland}

\author{Antoni Szczurek\footnote{also at University of Rzesz\'ow, 
PL-35-959 Rzesz\'ow, Poland}
}
\email{Antoni.Szczurek@ifj.edu.pl} 
\affiliation{Institute of Nuclear
Physics, Polish Academy of Sciences, ul. Radzikowskiego 152, 
PL-31-342 Krak{\'o}w, Poland}
 
\begin{abstract}
We calculate the total cross section and transverse momentum distributions 
for the production of enigmatic $\chi_{c,1}(3872)$ (or X(3872)) assuming 
different scenarios: 
$c \bar c$ state and $D^{0*} {\bar D}^0 + D^0 {\bar D}^{0*}$ molecule.
The derivative of the $c \bar c$ wave function needed in the first
scenario is taken from a potential model calculations.
Compared to earlier calculation of molecular state we include not only 
single parton scattering (SPS) but also double parton scattering (DPS) 
contributions.  
The latter one seems to give smaller contribution than the SPS one.
The upper limit for the DPS production of 
$\chi_{c,1}(3872)$ is much below the CMS data.
We compare results of our calculations with existing experimental data 
of CMS, ATLAS and LHCb collaborations.
Reasonable cross sections can be obtained in either $c \bar c$
or molecular $D {\bar D}^*$ scenarios for $X(3872)$.
Also a hybrid scenario is not excluded.
\end{abstract}

%\pacs{12.38.Bx, 13.85.Ni, 14.40.Pq}
\maketitle

%----------------------------
\section{Introduction}
%----------------------------

The $X(3872)$ state was  discovered already some time ago by the Belle collaboration \cite{Belle_X}. Since then its existence has been confirmed in several other processes and numerous theoretical studies have been performed, see for example the review articles \cite{Lebed:2016hpi,Olsen:2017bmm,Brambilla:2019esw}.
There is at present agreement that the $X(3872)$ has the axial vector quantum numbers $J^{PC} = 1^{++}$, and accordingly the state is named as $\chi_{c1}(3872)$ \cite{PDG}.

The internal structure of $X(3872)$ stays rather enigmatic. While its quantum numbers are not exotic --
it could indeed be a quarkonium $c \bar c$ state, e.g. a radial
excitation of the $\chi_{c1}$, there are strong arguments for a non-$c
\bar c$ scenario, manifested e.g. by the violation of isospin in its
decays \cite{PDG}.

More importantly, the mass of $X$ is very close to the $D {\bar D}^*$ threshold.
It is therefore rather popular to consider $X(3872)$ as very weakly bound state of the $D {\bar D}^*$ system - a hadronic
molecule, see the review \cite{Guo:2017jvc}. 
A tetraquark scenario was considered in \cite{MPPR2005}. 
The $c \bar c$ quarkonium scenario, where $X(3872)$ is the $\chi_{c1}(2P)$ state has been advocated in \cite{Achasov:2015oia}. 
Other approaches treat the $X(3872)$ as a $c \bar c$ bound state in the meson-meson continuum taking into account coupling of $c \bar c$ and meson-meson channels \cite{Cincioglu:2016fkm,CRB2013,Giacosa:2019zxw}. The possible mixture of quarkonium and molecule/virtual state is considered in \cite{Kang:2016jxw} and found to be consistent with current data.

Recently, the transverse momentum distributions of $X(3872)$ were measured at the LHC \cite{CMS_X,ATLAS_X,LHCb_X}
by the CMS, ATLAS and LHCb collaborations.

There is a debate in the literature \cite{Bignamini:2009sk, Artoisenet:2009wk, Albaladejo:2017blx,Wang:2017gay}, whether the rather large production rate at large $p_T$ allows one to exclude the molecular scenario. With few exceptions \cite{Butenschoen:2013pxa,MHC2013,Ilten:2021jcb}
the discussion in the literature is limited to estimates of orders of magnitude.

In this paper we shall consider two 
scenarios of prompt $X(3872)$
production, which are not mutually exclusive and in fact both can contribute depending
on the structure of $X(3872)$. 
Both scenarios have in common that the production is initiated by the production of a $c \bar c$ pair in a hard process.

In the first scenario we shall consider that 
the $X(3872)$ is a pure $c \bar c$ state,
the first radial excitation of $\chi_{c1}$.
The corresponding wave function and its derivative were calculated in potential models e.g. in \cite{EQ2019}. 

In the second scenario, where the $X(3872)$ is treated as a weakly bound $s$-wave state in the $D \bar D^* + D^* \bar D$-system, we exploit the connection between the low-energy scattering amplitude in the continuum and at the bound state pole below threshold, well known from effective range theory. In this work we follow \cite{Artoisenet:2009wk}
and give an estimate of a $p_T$-dependent upper bound for $X(3872)$ production in the molecule scenario.

As far as the hard production mechanism is concerned, we employ the $k_T$-factorization 
framework \cite{Catani:1990eg,Levin:1991ry,Collins:1991ty}. 
For the quarkonium scenario, the dominant mechanism of $C = +$1 is probably color singlet two virtual gluon fusion. The production of $\chi_{c0}, \chi_{c1}, \chi_{c2}$ production 
was considered e.g. in \cite{Samara,Baranov:2015yea,CS2018}.
Recently we have shown that the transverse momentum distribution 
of $\eta_c$ measured by the LHCb \cite{LHCb_etac} can be nicely 
described as $g^* g^* \to \eta_c$ fusion within $k_T$-factorization
approach \cite{BPSS2020}. 

For the molecular scenario, in addition to the single-parton scattering (SPS) mechanism of fusion of two off-shell gluons $g^* g^* \to c \bar c$, we 
will also consider production through the double-parton scattering (DPS) mode. 

%-----------------------
\section{Formalism}
%-----------------------

\begin{figure}
{
    \centering
    \includegraphics[width=.4\textwidth]{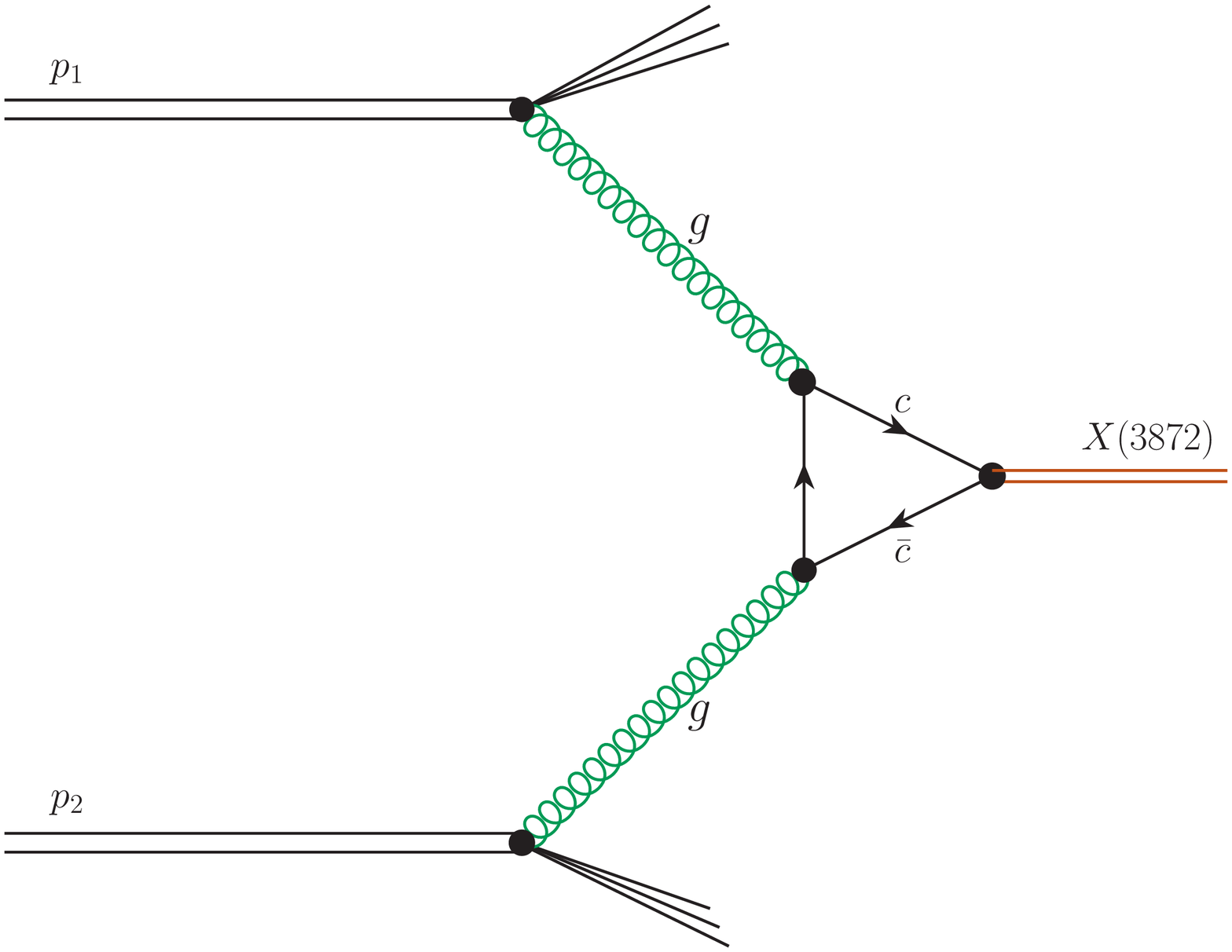}
    \includegraphics[width=.4\textwidth]{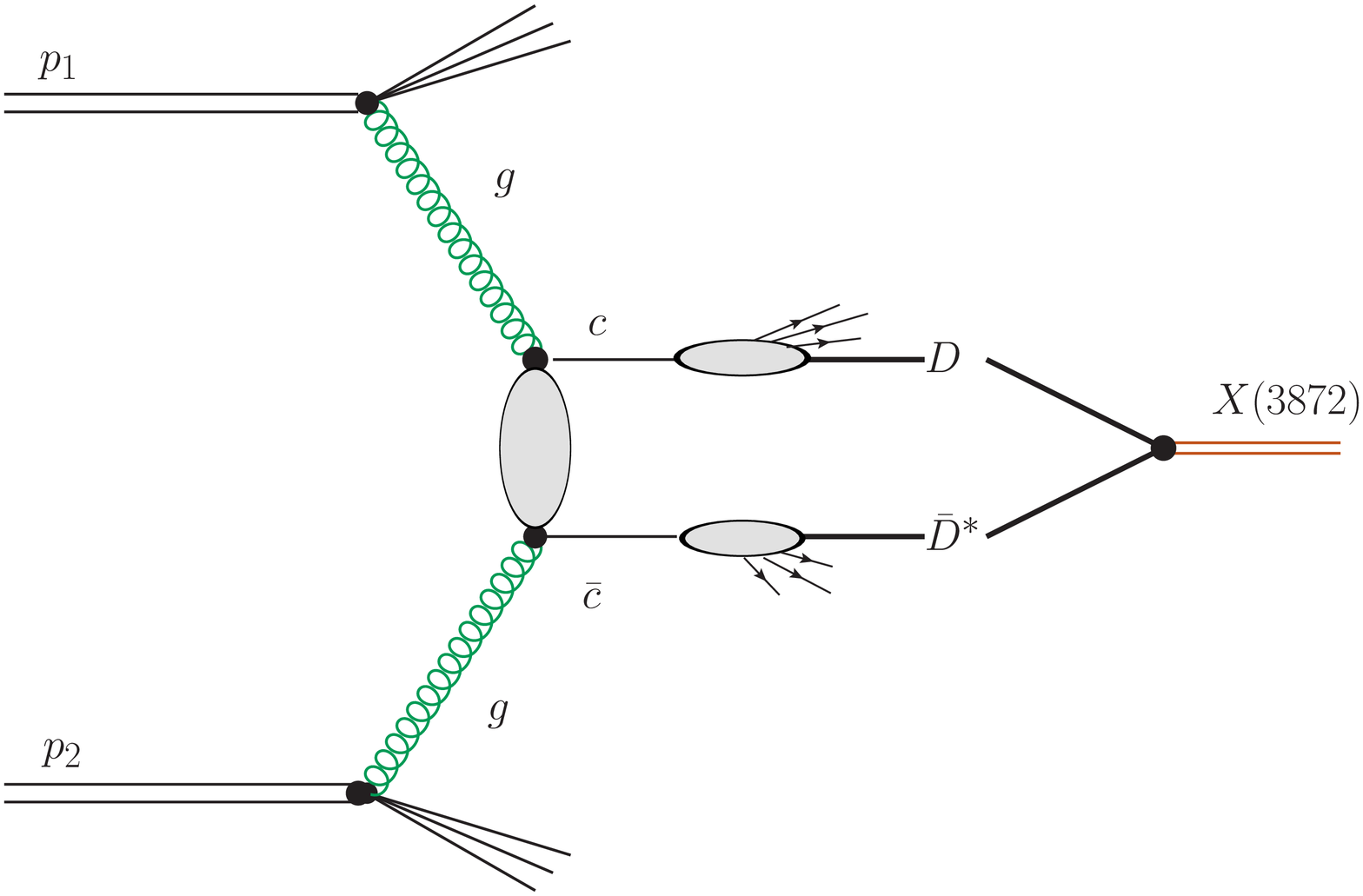}
    }
   \caption{Generic diagrams for the inclusive process of 
$X(3872)$ production in proton-proton scattering via two gluons fusion. }
    \label{fig:diagrams_X}
\end{figure}

In Fig.\ref{fig:diagrams_X} we show two generic Feynman diagrams for 
$X(3872)$ quarkonium production in proton-proton collision via 
gluon-gluon fusion: for the quarkonium (left) and molecule (right).
These diagrams illustrate the situation adequate for 
the $k_T$-factorization calculations used in the present paper.

%--------------------------------------------------------------------------
%---------------------------------------------------------------------------

The inclusive cross section for $X(3872)$-production via 
the $2 \to 1$ gluon-gluon fusion mode is obtained from
%%%%
\begin{eqnarray}
{d \sigma \over dy d^2\bp} &=& \int {d^2 \bq_1 \over \pi \bq_1^2} 
{\cal{F}}(x_1,\bq_1^2,\mu_F^2) \int {d^2 \bq_2 \over \pi \bq_2^2}
{\cal{F}}(x_2,\bq_2^2,\mu_F^2) \, \delta^{(2)} (\bq_1 + \bq_2 - \bp ) 
\,
\nonumber \\
&\times&{\pi
  \over (x_1 x_2 s)^2} \overline{|{\cal{M}}_{g^* g^* \to X(3872)}|^2} \,  .
\label{eq:cross_section_2}
\end{eqnarray}
%%%%
Here the matrix element squared for the fusion of two off-shell gluons into the $^3 P_1$ color singlet $c \bar c$ charmonium is (see e.g. \cite{Samara,Cisek:2017ikn} for a derivation):
%%%%
\begin{eqnarray}
\overline{|{\cal{M}}_{g^* g^* \to X(3872)}|^2}  &=& (4 \pi \alpha_S)^2 \, {4 |R'(0)|^2 \over \pi M_X^3} {\bq_1^2 \bq_2^2 \over (M_X^2 + \bq_1^2 + \bq_2^2)^4} \nonumber \\
&\times&
\Big( (\bq_1^2 + \bq_2^2)^2  \sin^2 \phi + M_X^2 ( \bq_1^2 + \bq_2^2 - 2 |\bq_1||\bq_2| \cos \phi )\Big) \, , 
\end{eqnarray}
%%%
where $\phi$ is the azimuthal angle between $\bq_1,\bq_2$.
The momentum fractions of gluons are fixed as 
$x_{1,2} = m_T \exp(\pm y) / \sqrt{s}$, where $m_T^2 = \bp^2 + M_X^2$.
The derivative of the radial quarkonium wave function at the origin is taken for the first radial $p$-wave excitation from Ref.\cite{EQ2019}, $|R'(0)|^2 = 0.1767 \, {\rm GeV}^5$.

The unintegrated gluon parton distribution functions (gluon uPDFs) are normalized such, that the collinear glue is obtained from
%%%%
\begin{eqnarray}
xg(x,\mu_F^2) = \int^{\mu_F^2} {d^2\bk \over \pi \bk^2} {\cal{F}}(x,\bk^2,\mu_F^2) \, .
\end{eqnarray}
%%%%
The hard scale is taken to be always $\mu_F = m_T$, the transverse mass of the $X(3872)$.
In order to estimate the production cross section for the molecule we also start from a hard production of a $c \bar c$-pair, which we then hadronize into a $D \bar D^* + h.c$ system using a prescription  given below.

The parton-level differential cross section for the 
$c \bar c$ production, formally at leading-order, reads:
\begin{eqnarray}\label{LO_kt-factorization} 
\frac{d \sigma(p p \to Q \bar Q \, X)}{d y_1 d y_2 d^2\bp_{1} d^2\bp_{2}} &=&
\int \frac{d^2 \bk_{1}}{\pi \bk^2_1}
{\cal F}(x_1,\bk_{1}^2,\mu_{F}^2)
\int \frac{d^2 \bk_{2}}{\pi \bk^2_2}
{\cal F}(x_2,\bk_{2}^2,\mu_{F}^2)
\nonumber \\
&&\times 
\delta^{(2)} \left( \bk_{1} + \bk_{2} - \bp_{1} - \bp_{2} \right) \
\frac{1}{16 \pi^2 (x_1 x_2 s)^2} \; \overline{ | {\cal M}^{\mathrm{off-shell}}_{g^* g^* \to c \bar c} |^2} \, .
\end{eqnarray}
where  
${\cal M}^{\mathrm{off-shell}}_{g^* g^* \to Q \bar Q}$ is the off-shell matrix element for the hard subprocess \cite{Catani:1990eg}, we use its implementation from \cite{maciula}.

Here, one keeps exact kinematics from the very beginning and
additional hard dynamics coming from transverse momenta of incident
partons. Explicit treatment of the transverse momenta makes the approach
very efficient in studies of correlation observables. The
two-dimensional Dirac delta function assures momentum conservation. The
gluon uPDFs must be evaluated at longitudinal momentum fractions:
\begin{eqnarray} 
x_1 &=& \frac{m_{T1}}{\sqrt{s}}\exp(+y_1) + \frac{m_{T2}}{\sqrt{s}}\exp(+y_2) \; , \\
x_2 &=& \frac{m_{T1}}{\sqrt{s}}\exp(-y_1) + \frac{m_{T2}}{\sqrt{s}}\exp(-y_2), 
\label{x1x2}
\end{eqnarray}
where $m_{Ti} = \sqrt{p_{Ti}^2 + m_c^2}$ is the quark/antiquark transverse mass.  

In the present analysis we employ the heavy $c$-quark approximation and assume that three-momenta in the $pp$-cm frame are equal:
\begin{equation}
\vec{p}_D = \vec{p}_c  \; .
\label{approximation}
\end{equation}
This approximation could be relaxed in future.
%but is difficult technically.
The hadronization is then included only via fragmentation branching fractions:
\begin{eqnarray}
&&P(c \to D^0) = P({\bar c} \to {\bar D}^0)= 0.56 \; , \\
&&P(c \to D^+) = P({\bar c} \to D^-) = 0.225 \; ,\\
&&P(c \to D^{*0}) = P({\bar c} \to {\bar D}^{*0}) = 0.236 \; , \\
&&P(c \to D^{*+}) = P({\bar c} \to {\bar D}^{*-}) = 0.236 \; . 
\label{hadronization}
\end{eqnarray}
The first number is from \cite{Lisovyi:2015uqa} while the other numbers 
are from \cite{ATLAS_D-mesons}.

The cross section for $c \bar c$ production are then multiplied by
\begin{eqnarray}
\frac{1}{2} [ P(c \to D^0) P({\bar c} \to {\bar D}^{*0})
            + P(c \to D^{*0}) P({\bar c} \to {\bar D}^0) ] 
= 0.1322
\label{multiplication_factor}
\end{eqnarray}
and, when comparing to experimental data by the relevant branching fraction i.e. by ${\rm BR}(X \to J/\psi \pi^+ \pi^-)$ for the case of CMS and LHCb, and by 
${\rm BR}(X \to J/\psi \pi^+ \pi^-) {\rm BR}(J/\psi \to \mu^+ \mu^-)$ for the ATLAS data. The branching fractions are taken from \cite{PDG}.
The factor $\frac{1}{2}$ is related to the factor $\frac{1}{\sqrt{2}}$
in the definition of the molecular wave function, %\ref{molecular_wave_function}.
\begin{equation}
|\ket{\Psi_{mol}} = \frac{1}{\sqrt{2}}\left( |\ket{D^0 {\bar D}^{*0}} +|\ket{D^{*0} {\bar
    D}^0} \right) \; .
\label{molecular_wave_function}
\end{equation}
In our calculation, we control the dependence on the relative momentum of quark and antiquark in the rest frame of the pair:
\begin{equation}
k_{rel} = \half \sqrt{M^2_{c \bar c} - 4 m_c^2} \; ,
\label{relative_momentum}
\end{equation}
where $M_{c \bar c}$ is invariant mass of the $c \bar c$ system 
and $m_c$ is the quark mass.
In order to obtain an upper bound
for the molecule production
cross section, we should integrate the $D \bar D^*$ cross section over the relative momentum $k^{D \bar D^*}_{rel}$ up
to a cutoff $k^{D \bar D}_{max}$  \cite{Artoisenet:2009wk}.
We will instead impose a cutoff $k_{max}$ on the relative momentum $k_{rel}$. Within our kinematics the latter will be similar to $k_{rel}^{D \bar D^*}$,
but somewhat larger.
In reality, for larger $\bp_{X} \approx \bp_{1} + \bp_{2}$, we have $k_{DD} < k_{rel}$.
We therefore estimate, that $k_{max} = 0.2 \, {\rm GeV}$ corresponds roughly to $k^{DD}_{max} \approx 0.13 \, \rm{GeV}$.
A better approximation would be to add simultaneous $c \to D$, 
${\bar c} \to {\bar D}$ fragmentation to our Monte Carlo code, which however means at least two more integrations.
We do not consider here any model of the $D {\bar D}^*$ wave function.

What is the appropriate choice for $k^{D \bar D^*}_{max}$ was a matter of discussion in the literature. In Ref. \cite{Bignamini:2009sk} it was suggested, that $k^{D \bar D^*}_{max}$ should be of the order of the binding momentum $k_X = \sqrt{2 \mu \varepsilon_X}$, where $\mu$ is the reduced mass of the $D \bar D^*$-system, and $\varepsilon_X$ is the binding energy of $X(3872)$.
This would lead to a very small value for $k^{D \bar D^*}_{max}$, similar to $k^{D \bar D^*}_{max} = 35 \, \rm{MeV}$ used in
\cite{Bignamini:2009sk}. 
However problems with this estimate have been pointed out in \cite{Artoisenet:2009wk,Albaladejo:2017blx,Wang:2017gay}. As has been argued in these works, the integral should be extended rather to a scale $k^{D \bar D^*}_{\max} \sim m_\pi$. In our choice of $k_{max}$, we follow this latter prescription.

In the following for illustration we shall therefore assume $k_{max} = 0.2 \, \rm{GeV}$.
The calculation for the SPS molecular scenario is done using the VEGAS algorithm for Monte-Carlo integration \cite{Lepage:1977sw}.

\begin{figure}
{
    \centering    \includegraphics[width=.4\textwidth]{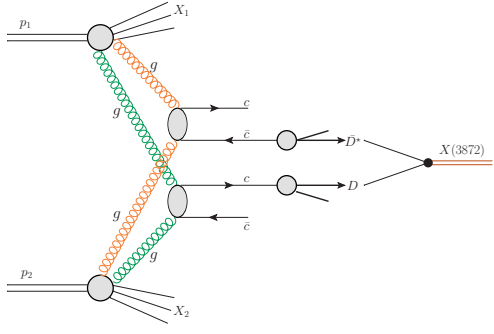}
    }
   \caption{A generic diagram for the inclusive process of 
$X(3872)$ production in proton-proton scattering via the double parton scattering mode. }
    \label{fig:diagram_X_DPS}
\end{figure}

We also include double parton scattering contributions (see Fig.\ref{fig:diagram_X_DPS}).
%%%
%%%
The corresponding cross section is calculated in the so-called
factorized ansatz as:
\begin{equation}
\Delta \sigma =
\frac{1}{2 \sigma_{\rm eff}}
\int \frac{d \sigma_{c \bar c}}{d y_1 d^2 \bp_{1}}
     \frac{d \sigma_{c \bar c}}{d y_2 d^2 \bp_{2}} \; \;
dy_1 d^2 \bp_{1} dy_2 d^2 \bp_{2} \Big|_{k_{rel} < k_{max}}  \; .
\label{DPS_contribution}
\end{equation}
Above the differential distributions of the first and second parton
scattering $\frac{d \sigma}{d y_i d^2 \bp_{i}}$ are calculated
in the $k_T$-factorization approach as explained above.
In the following we take $\sigma_{\rm eff}$ = 15 mb as in 
\cite{Maciula:2013kd}.
The differential distributions (in $p_{T}$ of the $X(3872)$ or 
$y_{\rm diff} = y_1 - y_2$, etc.) are obtained by binning in 
the appropriate variable. 
We include all possible fusion combinations leading to $X(3872)$:
\begin{eqnarray}
c_1 \to D^0, {\bar c}_2 \to {\bar D}^{*0}  \; , \\
c_1 \to D^{*0}, {\bar c}_2 \to {\bar D}^0  \; , \\
{\bar c}_1 \to {\bar D}^0, c_2 \to D^{*0}  \; , \\
{\bar c}_1 \to {\bar D}^{*0}, c_2 \to D^0  \; .
\label{fragmentation_factor}
\end{eqnarray}
This leads to the multiplication factor two times bigger than for the
SPS contribution (see Eq.(\ref{multiplication_factor})).

%--------------------
\section{Results}
%--------------------

In this section we shall show our results for recent CMS \cite{CMS_X}
ATLAS \cite{ATLAS_X} and LHCb \cite{LHCb_X} data. The CMS data is for 
$\sqrt{s}$ = 7 TeV and -1.2 $ < y_X < $ 1.2, 
the ATLAS data for $\sqrt{s} =$ 8 TeV, -0.75 $ < y_X < $ 0.75
and the LHCb data for $\sqrt{s} =$ 13 TeV, \,  2 $ < y_X < $ 4.5.
In all cases the $X(3872)$ was measured in the $J/\psi \pi^+ \pi^-$ channel.
We have used a number of different unintegrated gluon distributions,
firstly a distribution obtained from the solution of a BFKL equation 
with kinematic constraints by Kutak and Sta\'sto denoted KS
\cite{Kutak:2004ym}, secondly a  gluon uPDF obtained from 
a modified  Kimber-Martin-Ryskin (KMR) procedure 
\cite{Kimber:2001sc,Martin:2009ii} based on Durham group 
collinear PDFs \cite{Harland-Lang:2014zoa}. Finally we 
also employ a gluon uPDF obtained by Hautmann and Jung 
\cite{Hautmann:2013tba} from a description of precise HERA data 
on deep inelastic structure function by a solution of the CCFM evolution equations.

%---------------------------------
\subsection{$c \bar c$ state}
\begin{figure}
\includegraphics[width=0.8\textwidth]{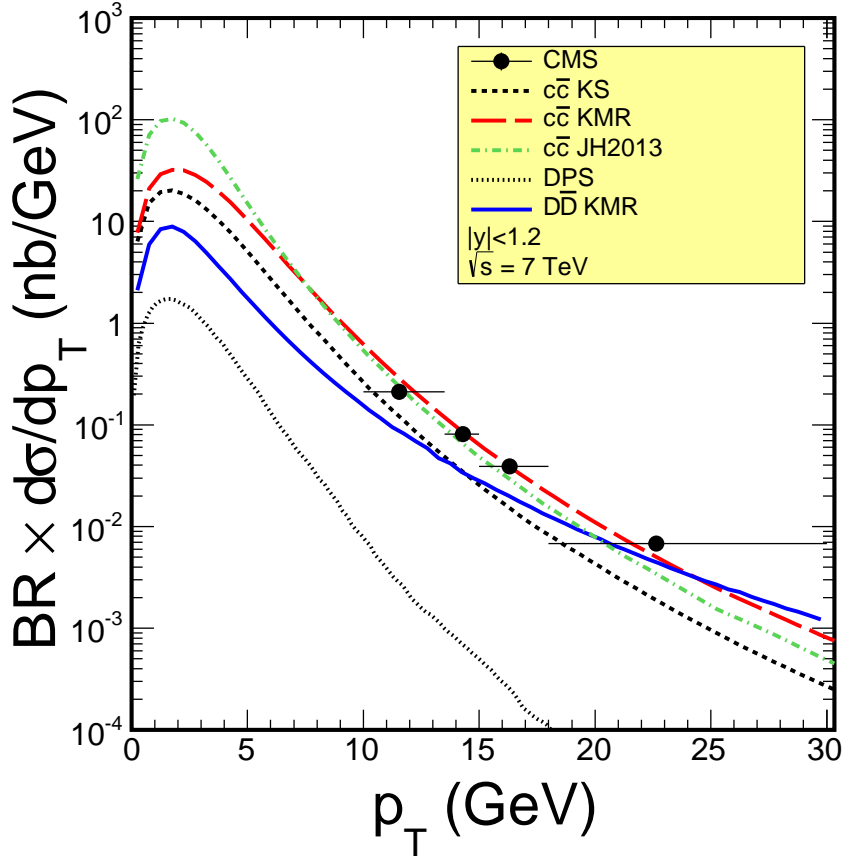}
\caption{Transverse momentum distribution of $X(3872)$ for
the CMS experiment.
Shown are results for 3 different gluon uPDFs.
Here BR = 0.038.
The upper limit for the SPS molecular scenario is shown as the solid line.
We also show corresponding distribution for the DPS mechanism (dotted line).
}
\label{fig:ccbar_state_CMS}
\end{figure}

\begin{figure}
\includegraphics[width=0.8\textwidth]{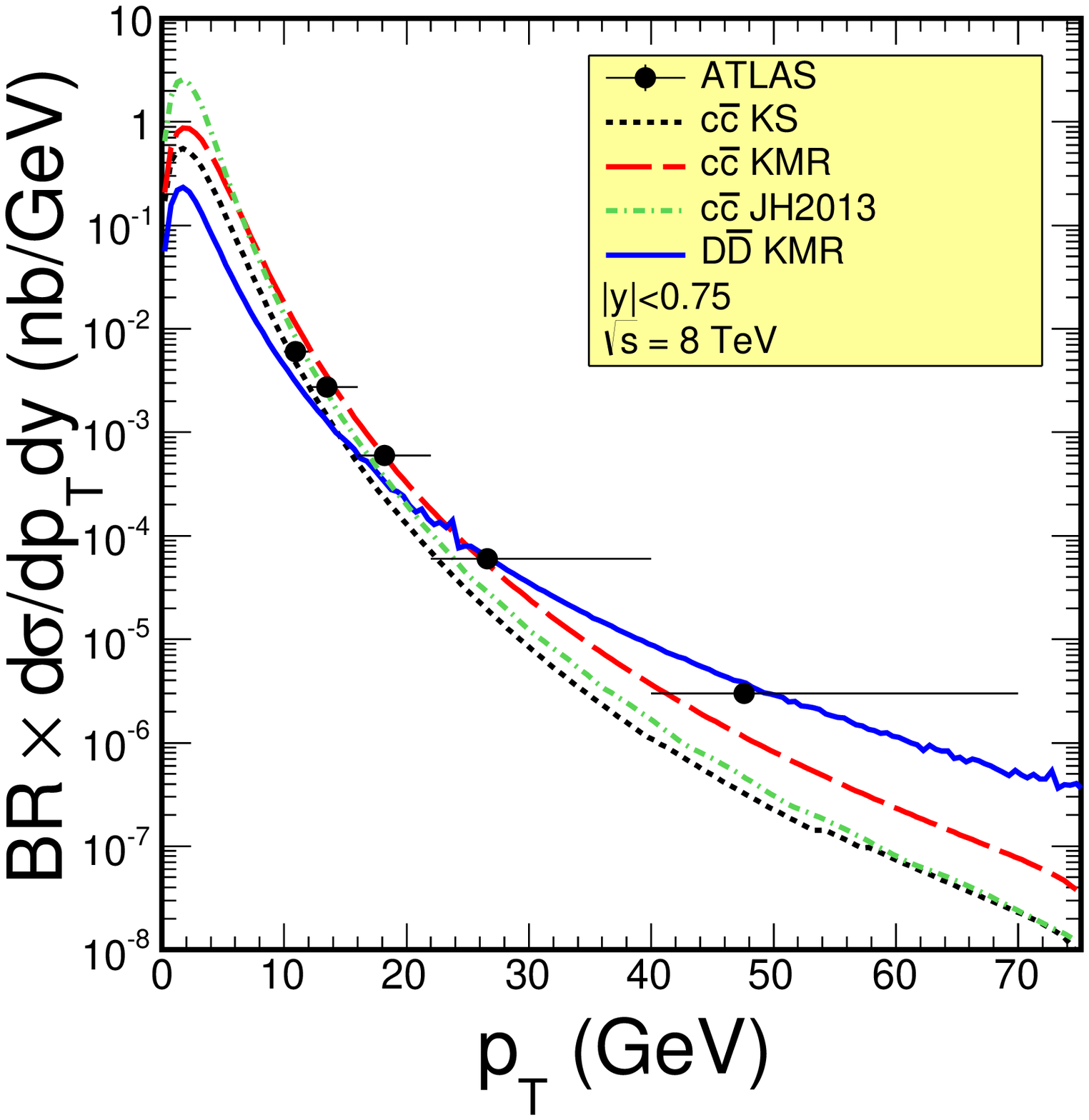}
\caption{Transverse momentum distribution of $X(3872)$ for
the ATLAS experiment.
Shown are results for 3 different gluon uPDFs. Here ${\rm BR} = 0.038 \cdot 0.0596$.
The upper limit for the SPS molecular scenario is shown as the solid line.
}
\label{fig:ccbar_state_ATLAS}
\end{figure}

\begin{figure}
\includegraphics[width=0.8\textwidth]{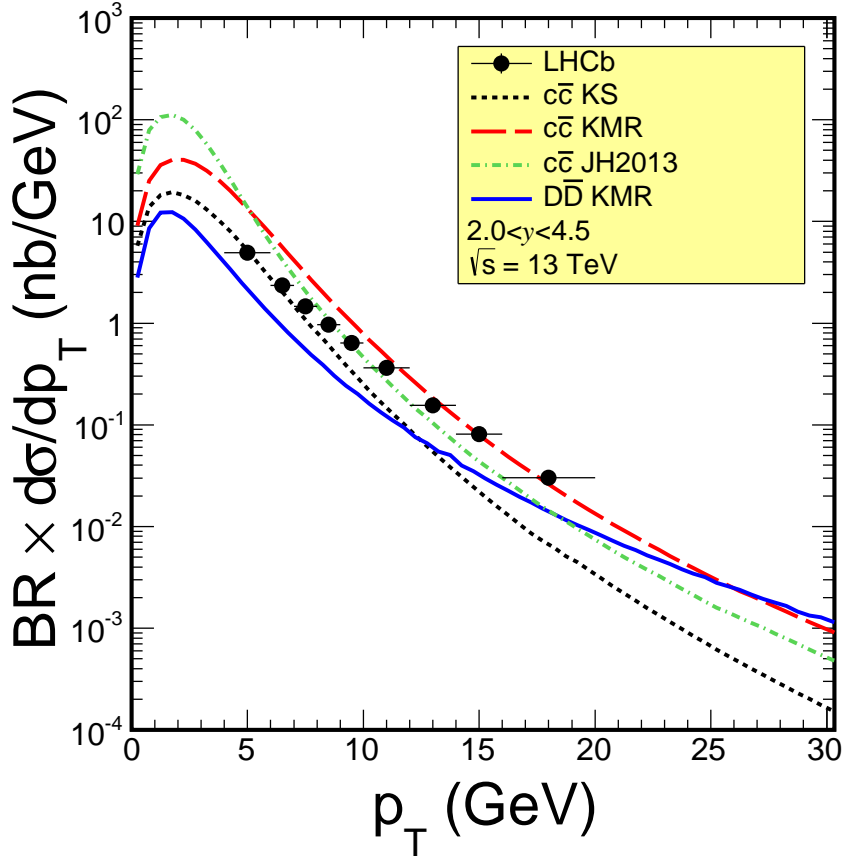}
\caption{Transverse momentum distribution of $X(3872)$ for
the LHCb experiment.
Shown are results for 3 different gluon uPDFs.
Here BR = 0.038.
The upper limit for the SPS molecular scenario is shown as the solid line.
}
\label{fig:ccbar_state_LHCb}
\end{figure}

%---------------------------------
Here we show our results for $\chi_{c,1}(3872)$ production treated
as a pure $c \bar c$ state.
In Figs.\ref{fig:ccbar_state_CMS},
\ref{fig:ccbar_state_ATLAS},\ref{fig:ccbar_state_LHCb}
we show the transverse momentum distribution of the $X(3872)$ together
with the data from the CMS, ATLAS and LHCb experimental data.
A surprisingly good description is obtained with different gluon uPDFs
specified in the figure legend without any free parameters.
It is worth to mention approximately good slope of the $p_{TX}$
distributions which is due to effective inclusion of higher-order
corrections. The corresponding result within the collinear leading-order
approximation would be equal to zero!
A slightly different slope is obtained for the molecular scenario (solid line)
discussed in detail somewhat below.
%
%-------------------------------------------------------------------
%-------------------------------------------------------------------
%

%-------------------------------------------------------------------
%-------------------------------------------------------------------
%

%-------------------------------------------------------------------
%-------------------------------------------------------------------
%
%--------------------------------------
\subsection{Molecular picture}
%---------------------------------------

\begin{figure}
\includegraphics[width=.45\textwidth]{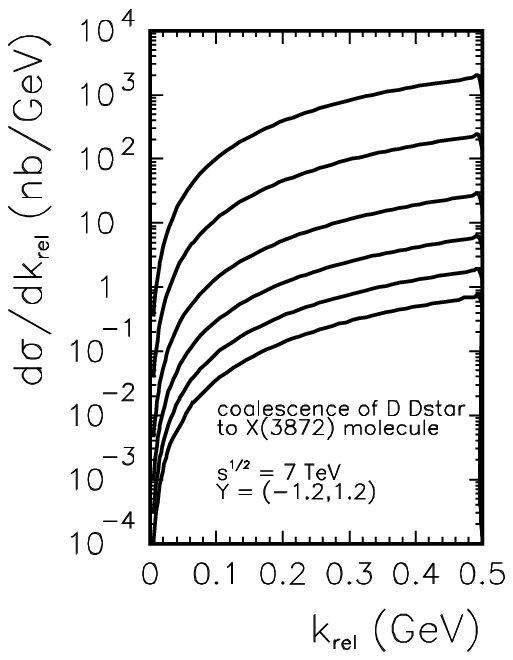}
\includegraphics[width=.45\textwidth]{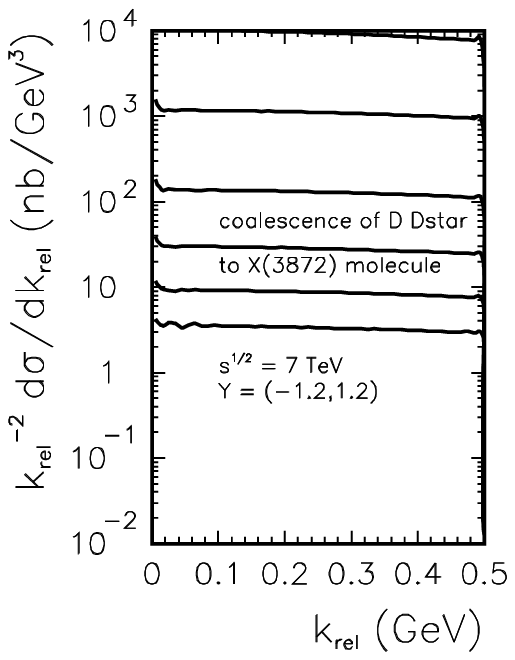}
\caption{Distribution in $k_{rel}$ for different windows of 
$p_{t,c \bar c} = p_{t,X}$ (left panel) for the CMS kinematics 
as specified in the main text. In the right panel we show the cross
sections divided  by $k_{rel}^2$.
In these calculations the KMR UGDF with the MMHT NLO collinear gluon 
distribution was used. Branching fractions are included here.
}
\label{fig:dsig_dkrel}
\end{figure}

Here we show our predictions for $\chi_{c,1}(3872)$ production treated
as the $\bar D^{*0} D^{0} + h.c.$ molecule.
Then the $\chi_{c,1}(3872)$ can be produced if $k_{rel}$ is small, and an estimate (or upper bound) for its production cross section can be obtained from the continuum  cross section at small $k_{rel}$.
In Fig.\ref{fig:dsig_dkrel} we show the distribution in this variable
for different windows of $|\bp_{c \bar c}|$ (0,5 GeV), (5,10 GeV),
(10,15) GeV, (15,20) GeV, (20,25) GeV, (25,30) GeV.

Similar shapes in $k_{rel}$ are obtained for the different windows
of $p_{T c \bar c}$.

%-------------------------------------------------------------
%-------------------------------------------------------------

To visualize this better we show in the right panel of Fig.\ref{fig:dsig_dkrel}
 $k_{rel}^{-2} d \sigma/d k_{rel}$.
As expected, the so-obtained distributions closely follow  phase-space, 
and are almost flat in a broad range of $k_{rel}$.
 Therefore, the cross section has essentially the phase-space behaviour
%%%%
\begin{eqnarray}
d \sigma \propto k_{rel}^2 dk_{rel} \, ,
\label{eq: k_rel-dependence}
\end{eqnarray}
%%%%
which implies the strong, cubic dependence $\propto k_{max}^3$ on the upper limit $k_{max}$ of the $k_{rel}$-integration.

The calculation in the whole phase space 
$p_{T1} \in (0,20) \, \rm{GeV}$ and $p_{T2} \in (0,20)\, \rm{GeV}$
leads to fluctuations at large $p_{Tc \bar c} > 20 \, \rm{GeV}$.
This can be understood as due to steep dependence of the cross
section on $p_{T1}, p_{T2}, y_1, y_2, \phi$.
%There steep dependence in the $p_{1t} \otimes p_{2t}$ space. 
Only a narrow range in the $p_{T1} \otimes p_{T2}$ space with $p_{T1} \approx p_{T2}$ fulfills 
the condition $k_{rel} < k_{max} = 0.2 \, \rm{GeV}$. 

 We remind the reader that this value of $k_{max}$ is imposed on the $c \bar c$ final state, and would correspond to a smaller $k_{max}^{DD^*} \sim 0.14 \, \rm{GeV}$ for the $D \bar D^*$ mesons. Due to the behaviour shown in Eq.(\ref{eq: k_rel-dependence}), the cross section for $k_{max} = 0.14 \, \rm{GeV}$ or $k_{max} = 0.1 \, \rm{GeV}$ imposed on the $c \bar c$ state would go down by a factor three or eight, repsectively.

\begin{figure}
\includegraphics[width=8cm]{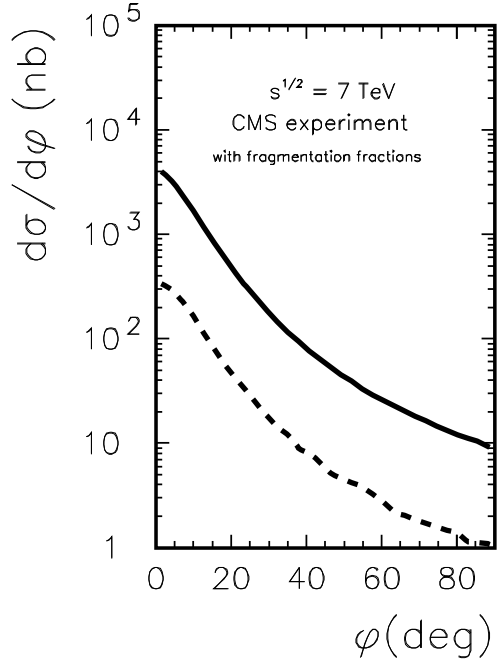}
\caption{Azimuthal correlations between $c \bar c$ that fullfill
the condition $k_{rel} <$ 0.2 GeV. Here the CMS cuts were imposed.
We show contribution of SPS (solid line) and DPS (dashed line).
}
\label{fig:dsig_dphi_CMS}
\end{figure}

The distribution in relative azimuthal angle between $c \bar c$
in the $pp$-frame with the cut  $k_{rel} < 0.2 \, \rm{GeV}$ is shown in Fig.\ref{fig:dsig_dphi_CMS}.
This is rather a steep distribution around $\phi$ = 0$^o$.
This is not a typical region of the phase space.
Recall, that in the leading-order collinear approach $\phi$ = 180$^o$. It is obvious that the region of $\phi \approx$ 0 cannot be obtained easily in the collinear approach.
As discussed in \cite{maciula} the $k_T$-factorization approach
gives a relatively good description of $D^0 {\bar D}^0$ correlations 
in this region of the phase space.

%------------------------------------------------------------------
%------------------------------------------------------------------

Having understood the kinematics of the $X(3872)$ production we can 
improve the description of transverse momentum distribution of
$X(3872)$.
The rather strong correlation $p_{T1} \approx p_{T2}$ allows to perform 
VEGAS calculation simultaneously limiting to $p_{T1}, p_{T2} > p_{T min}$.
We have also verified that
\begin{equation}
p_{T1}+p_{T2} \approx p_{T c \bar c} = |\bp_{1} + \bp_{2}|   \; .
\label{p1tplusp2t}
\end{equation}
Therefore imposing lower limits on 
$p_{T1} > p_{T min}$ and $p_{T2} > p_{T min}$ means
also lower limit on $p_{Tc \bar c} > 2 p_{T min}$.
The solid line in Fig.\ref{fig:ccbar_state_CMS} shows result of 
such a calculation. The fluctuations are gone.
Combining the two calculations at say $p_{T c \bar c}  = 15 \, \rm{GeV}$ gives
a smooth result everywhere.

%-----------------------------------
\subsection{Hybrid approach}
%-----------------------------------

\begin{figure}
\includegraphics[width=0.45\textwidth]{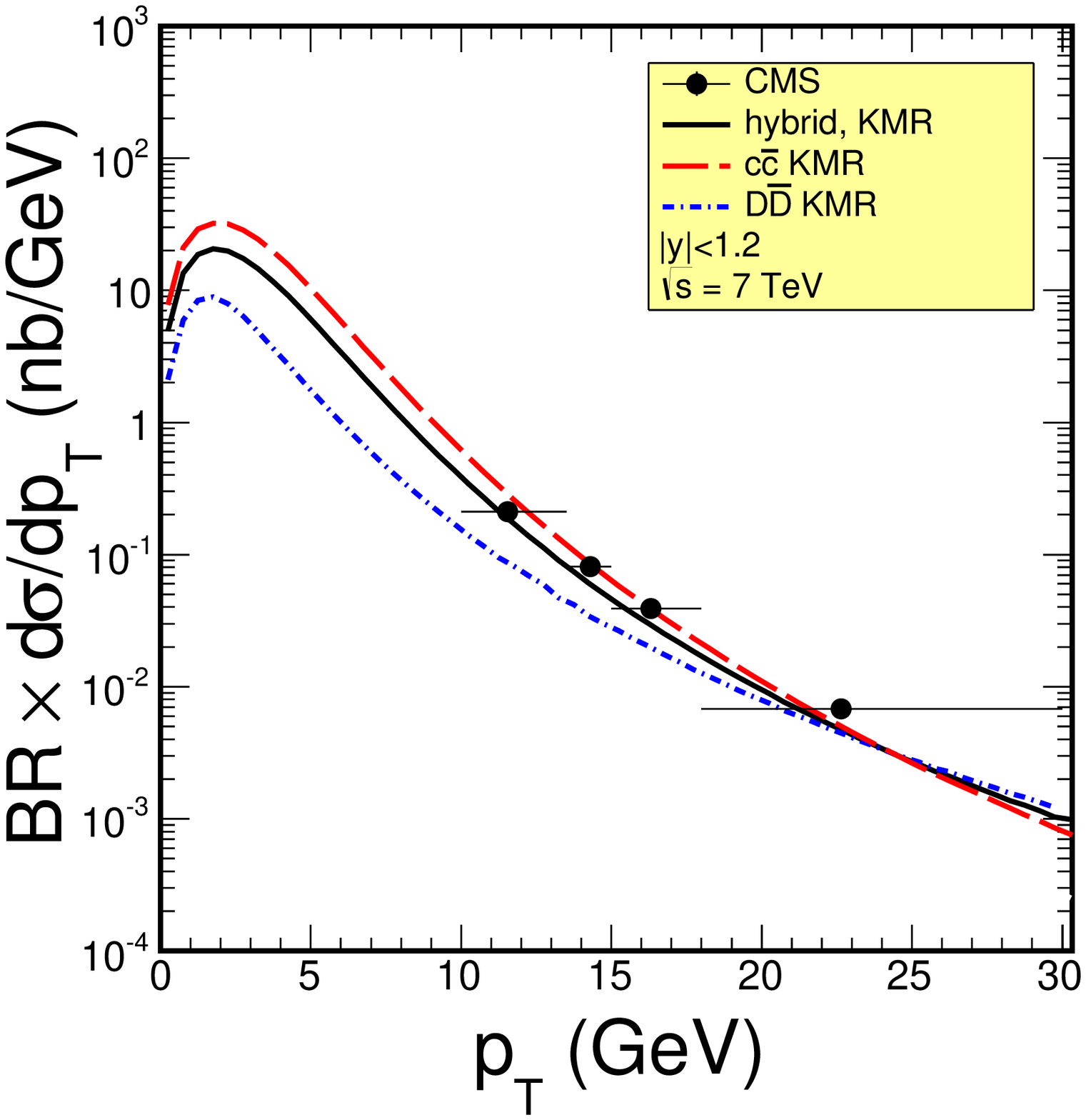}
\includegraphics[width=0.45\textwidth]{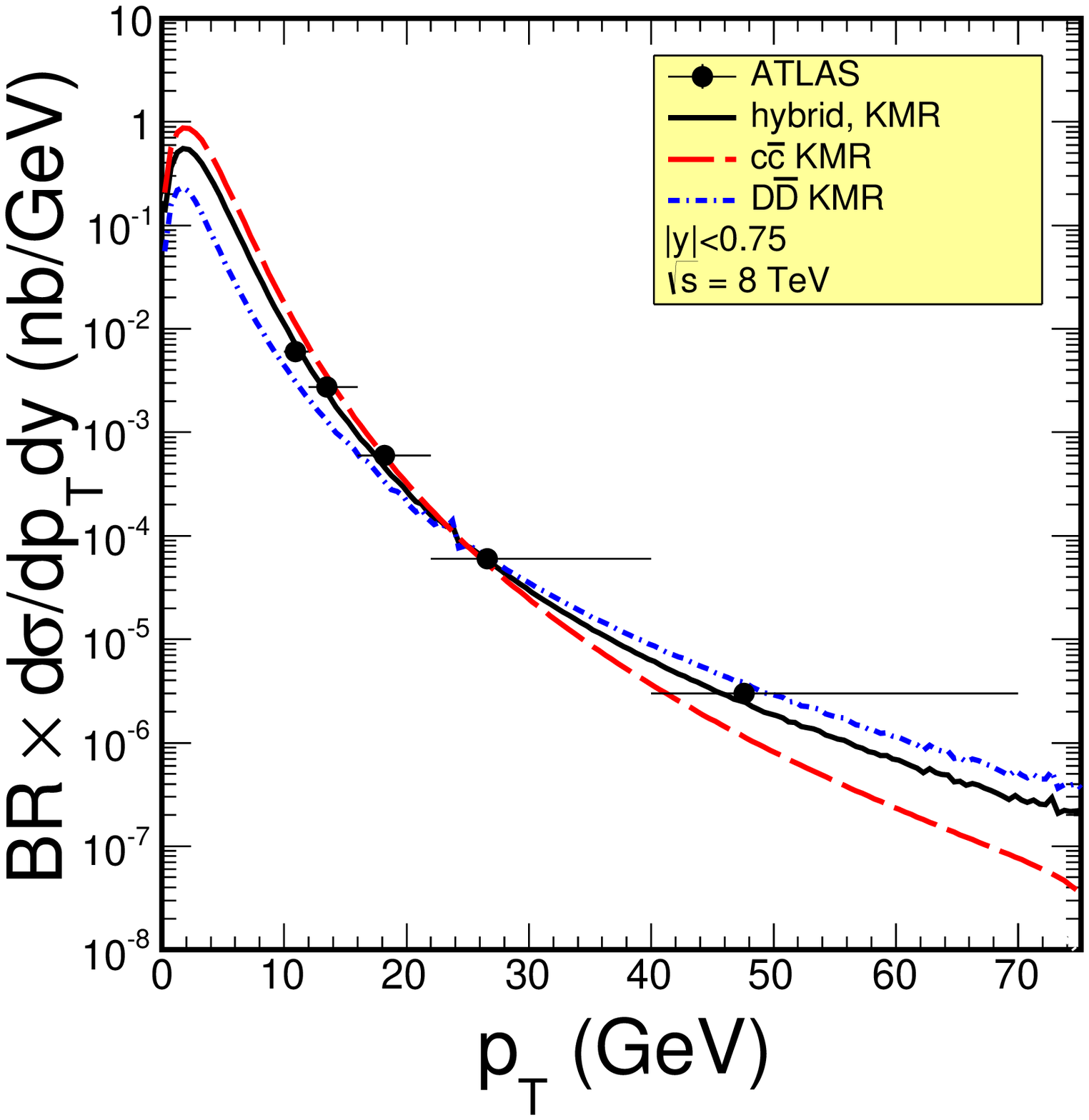}
\includegraphics[width=0.45\textwidth]{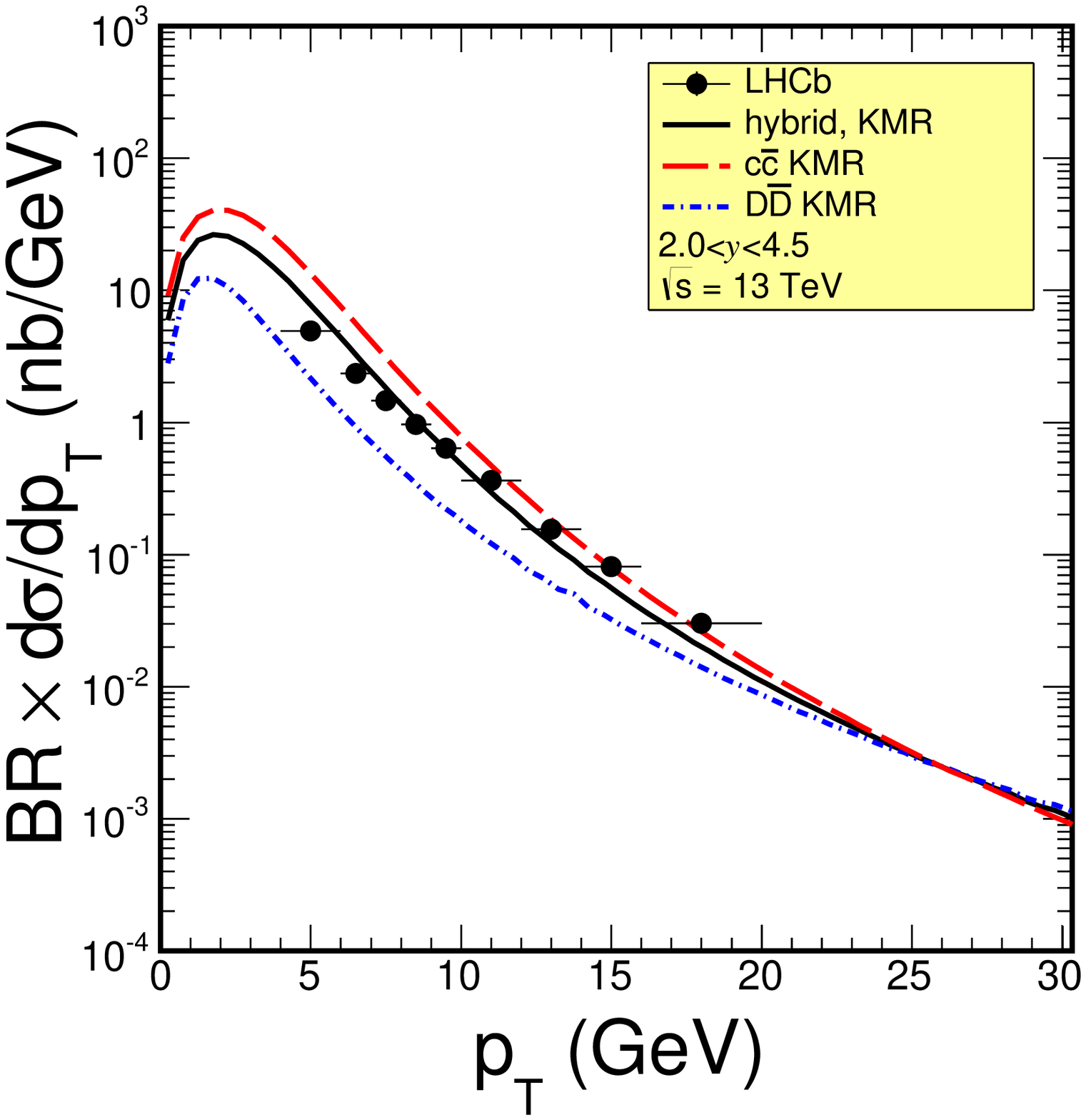}
\caption{Transverse momentum distribution of $X(3872)$ for
the CMS, ATLAS and LHCb experiments.
Shown are results for the KMR UGDF. Here BR = 0.038 for CMS and LHCb, and ${\rm BR} = 0.038 \cdot 0.0596$ for ATLAS.
We show results for different combinations of $\alpha$ and $\beta$ as
specified in the figure legend.
}
\label{fig:different_mixtures}
\end{figure}

Still another option is to assume that $X$ is the combination:
\begin{equation}
|\ket{X(3782)} = \alpha |\ket{c \bar c}
        + \frac{\beta}{\sqrt{2}} \left( |\ket{D {\bar D}^*} + |\ket{{\bar D} D^*} \right ) \; .
\label{hybrid}
\end{equation}
Such a state could be called hybrid.
Since the cross section for production of conventional and molecular 
$X(3872)$ are rather similar,
it is difficult to select one combination of $\alpha$ and $\beta$.
For example in Fig.\ref{fig:different_mixtures} we show also such 
a result for $\alpha^2 = \beta^2$ = 0.5 (dash-dotted line).
The corresponding result is similar to the result for $\alpha^2$ = 1, $\beta^2$ = 0 
(pure $c \bar c$ state) and $\alpha^2$ = 0, $\beta^2$ = 1 (pure molecular
state) calculated for the KMR UGDF.
The fifty-fifty mixture gives quite good representation of all the
experimental data.

%----------------------------------------------------------------------
%-------------------------------------------------------------------------

In summary, it is very difficult at present to definitely conclude what
is the mechanism of the $X(3872)$ production.
The same is true for the wave function of the $X(3872)$ meson 
(see Eq.(\ref{hybrid})).
At present we cannot answer the question whether $X(3872)$ is of 
conventional, molecular or hybrid type. More work is definitely needed in future.

%--------------------------
\section{Conclusions}
%--------------------------

We have performed the calculation of $X(3872)$ production
at the LHC energies. We have performed two independent calculations:
one within nonrelativistic QCD approach assuming pure $c \bar c$ state and
second assuming a coalescence of $D$ and ${\bar D}^{*}$
or $\bar D$ and $D^*$, consistent with molecular state assumption.
The first calculation requires usage of derivative of the $c \bar c$
wave function. In the present analysis we have used the wave function 
obtained in \cite{EQ2019}.
The resulting cross section was calculated within the $k_T$-factorization
approach with a few unintegrated gluon distributions.
In the second approach first the hard production of a $c \bar c$ pair is calculated.
Next a simple hadronization is performed giving a correlation 
distribution of $D$ and $D^*$ mesons.
Imposing limitations (upper limit) on relative momenta of $D$ and 
${\bar D}^*$ 
we get a $p_T$-dependent upper limit of the cross section for $D$-${\bar D}^*$ or ${\bar D}$-$D^*$ fusion 
(coalescence).
We compare, for the first time, to all available experimental data on the $p_T$-dependent cross section.
The both, quite different, scenarios give the cross section of the right order of magnitude, very similar to the CMS, ATLAS and LHCb 
experimental data.

Also a mixture of both mechanism leads to correct description of the 
CMS, ATLAS and LHCb data.
Our study within $k_T$-factorization approach shows that all the options
are in the game.
%--------------------
\acknowledgments
%--------------------
This work is partially supported by
the Polish National Science Centre under Grant No. 2018/31/B/ST2/03537
and by the Center for Innovation and Transfer of Natural Sciences 
and Engineering Knowledge in Rzesz\'ow (Poland).

%-------------------------------------------------------------------------------------

%-------------------------------------------------------------------------------------

\end{document}